\documentclass[fleqn,usenatbib]{mnras}

\usepackage{newtxtext,newtxmath}

\newcommand{\flux}{\,erg\,s$^{-1}$\,cm$^{-2}$} %
\usepackage[T1]{fontenc}
\usepackage{ae,aecompl}

\usepackage{graphicx}	
\usepackage{amsmath}	
\usepackage{amssymb}	

\usepackage[symbol]{footmisc}
\usepackage{fp}
\usepackage{multirow}
\usepackage{footnote}
\usepackage{tabularx}

\title[GJ 1151's corona]{The corona of GJ 1151 in the context of star-planet interaction}

\author[G. Foster et al.]{
G. Foster$^{1, 2}$\thanks{E-mail: gfoster@aip.de}, K. Poppenhaeger$^{1, 2}$, J. D. Alvarado-G\'omez$^{1}$ and J.H.M.M. Schmitt$^{3}$
\\
$^{1}$Leibniz Institute for Astrophysics Potsdam (AIP), An der Sternwarte 16, 14482 Potsdam, Germany\\
$^{2}$Universit\"at Potsdam,  Institut f\"ur Physik und Astronomie,  Karl-Liebknecht-Stra{\ss}e 24/25, 14476 Potsdam, Germany\\
$^{3}$Universit\"at Hamburg, Hamburger Sternwarte, Gojenbergsweg 112, 21029 Hamburg, Germany}

\date{Accepted 2020 July 2. Received 2020 June 29; in original form 2020 May 11
}

\pubyear{2015}

\begin{document}
\label{firstpage}
\pagerange{\pageref{firstpage}--\pageref{lastpage}}
\maketitle

\begin{abstract}
The low-mass star GJ 1151 has been reported to display variable low-frequency radio emission, which has been interpreted as a signpost of coronal star-planet interactions with an unseen exoplanet. Here we report the first X-ray detection of GJ 1151's corona based on XMM-Newton data. We find that the star displays a small flare during the X-ray observation. Averaged over the observation, we detect the star with a low coronal temperature of 1.6~MK and an X-ray luminosity of $L_X = 5.5\times 10^{26}$\,erg/s. During the quiescent time periods excluding the flare, the star remains undetected with an upper limit of $L_{X,\,qui} \leq 3.7\times 10^{26}$\,erg/s. 
This is compatible with the coronal assumptions used in a recently published model for a star-planet interaction origin of the observed radio signals from this star.
\end{abstract}




\section{Introduction}

Star-planet interactions are suspected to be able to alter stellar magnetic activity in a variety of ways. Tidal interaction may influence the rotational evolution and therefore the magnetic activity level of a host star, similar to tidal synchronization in close stellar binaries, or may influence convection in the outer layers of the star \citep{Cuntz2000, Pont2009, Pillitteri2014}. Magnetic interaction is thought to be able to manifest itself through processes like reconnection of stellar and planetary field lines \citep{Cuntz2000, Shkolnik2005, Lanza2008}, suppression of the stellar wind by preventing stellar magnetic loops from opening up \citep{Cohen2010}, triggering of stellar flares near the sub-planetary point \citep{Lanza2018, Fischer2019}, or sub-Alfv\'enic interaction, similar to the interaction seen in the Jupiter-Io system \citep{Goldreich1969}. In cases such as the Jupiter-Io interaction, where a planetary body is an obstacle in the flow of the plasma, Alfv\'enic waves are generated subsequent to the flow. The waves propagate along the magnetic field generating radiative energy, causing heating of the plasma \citep{Gosling1982, Saur2013, Strugarek2014, Turnpenney2018}.

Observational studies have reported some hints for tidal and magnetic interactions \citep{Shkolnik2005, Pont2009, Kashyap2008, Poppenhaeger2014, Maggio2015, Cauley2018}, but also caveats have been pointed out with respect to biases from planet-detection methods which may skew activity distributions in planet host samples \citep{Poppenhaeger2010, Miller2015}. The intrinsic variability of stellar activity on short and long time scales, such as flares or stellar activity cycles, makes an unambiguous attribution of stellar activity changes to a planetary origin challenging.

GJ~1151 is a low-mass star located in the solar neighbourhood; we list its basic physical parameters in Table~\ref{tab:info}. The star was observed to display variable radio emission \citep{Vedantham2020} with LOFAR \citep{vanHaarlem2013}. Several scenarios for a purely stellar origin of the radio emission were excluded, and \citet{Vedantham2020} concluded that sub-Alfv\'enic star-planet interaction with a so far undetected small planet in a close orbit is the most likely explanation for the observed radio signatures.

Here we report on the first X-ray detection of GJ~1151, and we present an analysis of the star's coronal properties in the context of star-planet interaction.

\section{Observations and data analysis}

\begin{table}\setlength{\tabcolsep}{15pt}
\caption{\label{tab:info} Fundamental physical parameters of the star GJ 1151.\newline $^a$\citet{GaiaSurvey2018} $^b$\citet{Skrutskie2006} $^c$\citet{Newton2017} $^d$\citet{Bailer-Jones2018}}
\centering
\begin{tabular}{l l }
\hline\hline
Parameter & Value \\
\hline 
Gaia DR2 ID & 786834302079285632$^a$  \\
2MASS ID &  J11505787+4822395$^b$    \\
$G$ (mag) & 11.694$^a$ \\
$J$ (mag) &  8.488$^b$ \\
$H$ (mag) &  7.952$^b$ \\
$K$ (mag) &  7.637$^b$ \\
mass & 0.167 $M_\odot$$^c$ \\
radius & 0.190 $R_\odot$$^c$ \\
distance & $8.036\pm 0.008$ pc$^d$ \\
\hline
\end{tabular}

\end{table}

\begin{figure*}
\includegraphics[width = 0.48\linewidth]{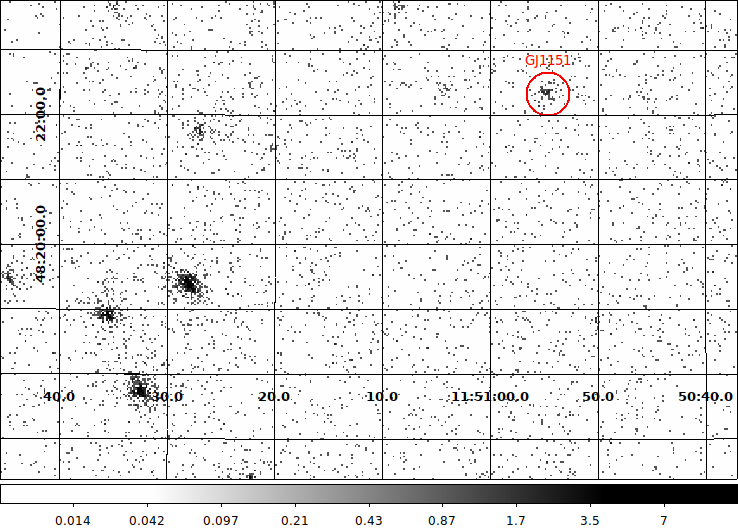}
\includegraphics[width = 0.48\linewidth]{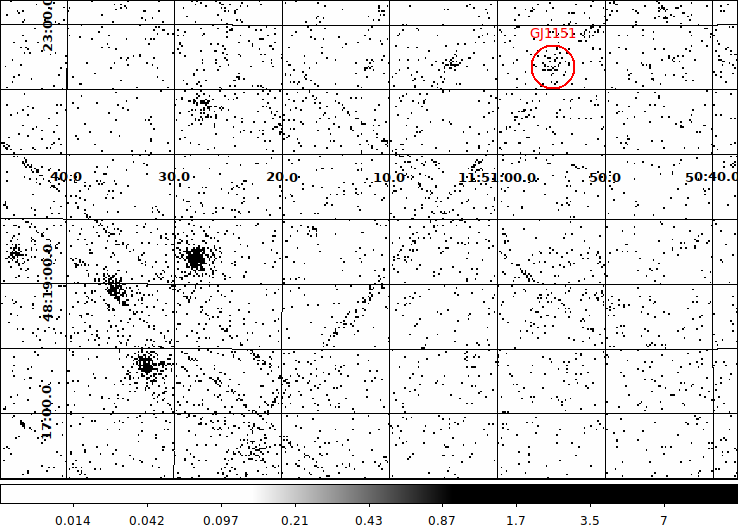}
\caption{X-ray image of GJ1151 observed in the 0.2-2 keV energy band with  \textit{XMM-Newton} on 1st November 2018. The left panel shows the combined image from the two MOS cameras, the right panel shows the image extracted from the PN detector where the target position was located on a chip edge. GJ1151 is marked by a circle with a $20^{\prime\prime}$ radius. }
\label{fig:xrayimages}
\end{figure*}

The star GJ~1151 was observed with \textit{XMM-Newton} on 1st November 2018 for 12 ks (ObsID 0820911301, PI J.\ Schmitt). The observations used the medium filter and full frame mode for all three CCD detectors (MOS1, MOS2, and PN). We analysed the data using XMM's SAS software version 18.0.0. We followed the standard data reduction steps outlined in the XMM SAS users guide\footnote{\url{https://xmm-tools.cosmos.esa.int/external/xmm_user_support/documentation/sas_usg/USG/}}, i.e.\ we filtered out bad-flag photon events and screened for times of high background using the full-chip high-energy count rates. Only the PN detector displayed significant time portions of high background. The MOS detectors displayed only such short time stretches of slightly elevated background that we opted to analyse the continuous MOS data, in order to facilitate a better analysis of the time variability of the source.

GJ~1151 is a star with a very long rotation period of 132 days \citep{Irwin2011} and low activity in the chromospheric H$\alpha$ line \citep{Newton2017}. Therefore its corona can be expected to emit mainly at soft X-ray energies below 2.0 keV. We therefore extracted X-ray images from the three CCD detectors in the 0.2-2.0 keV energy band, which we show in Fig.~\ref{fig:xrayimages}.

After taking into account the fast proper motion of the star, we placed a circular extraction region with $20^{\prime\prime}$ radius at GJ~1151's expected position during the epoch of the XMM-Newton observation, and defined a nearby source-free background region with a radius of $60^{\prime\prime}$. Unfortunately, GJ~1151's position fell on one of the PN detector's chip edges, so that only the data from the less sensitive MOS cameras could be used for further analysis. For the MOS detectors, we extracted light curves and CCD spectra following the standard procedures of the XMM SAS users guide.

\section{Results}

\subsection{An X-ray detection of GJ 1151 with XMM-Newton}

In Fig.~\ref{fig:xrayimages} we show X-ray images from XMM-Newton's MOS and PN cameras with the position of GJ1151 indicated. An excess is visible at the star's position in all cameras, but weaker in PN due to the closeness of GJ 1151's position to a detector chip edge.

To test whether GJ~1151 is significantly detected in X-rays, we used the Kraft-Burrows-Nousek (KBN) estimator \citep{Kraft1991} as implemented in the python astropy package \citep{Astropy2013, Astropy2018}. The KBN estimator takes as input the number of detected photons in a source detection region and the expected number of background photons in the same region, estimated from a larger source-free area; it assumes both numbers follow Poisson statistics, as is appropriate for X-ray photon counting. The KBN estimator marginalises over the possible background photons in the source detect region, and yields a confidence interval for the source counts in the source detection region.

In the 0.2-2 keV energy band, we find 43 and again 43 counts in the source extraction region for MOS1 and MOS2, respectively. For the same time intervals and energy band we find 112 and 82 counts in the nine times larger background extraction region (i.e.\ an expected background count rate of 12.4 and 9.1 per exposure in the source extraction region for MOS1 and MOS2, respectively). For both detectors individually the KBN estimator yields a detection at $>3\sigma$ level.

When combining the signal from both MOS detectors for smaller uncertainties, we therefore have 86 photons in the source region and 194 counts in the larger background region, collected over a total exposure time of $10.46+10.46=20.91$ks. We then derive a total number of background-subtracted source counts of $64.4^{+9.6}_{-8.9}$ with $1\sigma$ uncertainties for both MOS detectors co-added, again using the Kraft-Burrows-Nousek estimator. This translates to a background-subtracted count rate of 3.1 counts per ks for the combined MOS detectors for GJ 1151 in the 0.2-2 keV energy band.

We also checked if there is significant flux at energies above 2.0 keV, and found that there is no significant excess of counts in the energy bands of 2-5 keV or 2-10 keV. This is consistent with GJ~1151 being a soft X-ray emitter, as expected for a low-activity star.

\subsection{Temporal variability of GJ 1151's corona}

\begin{figure}
\includegraphics[width = 0.99\linewidth]{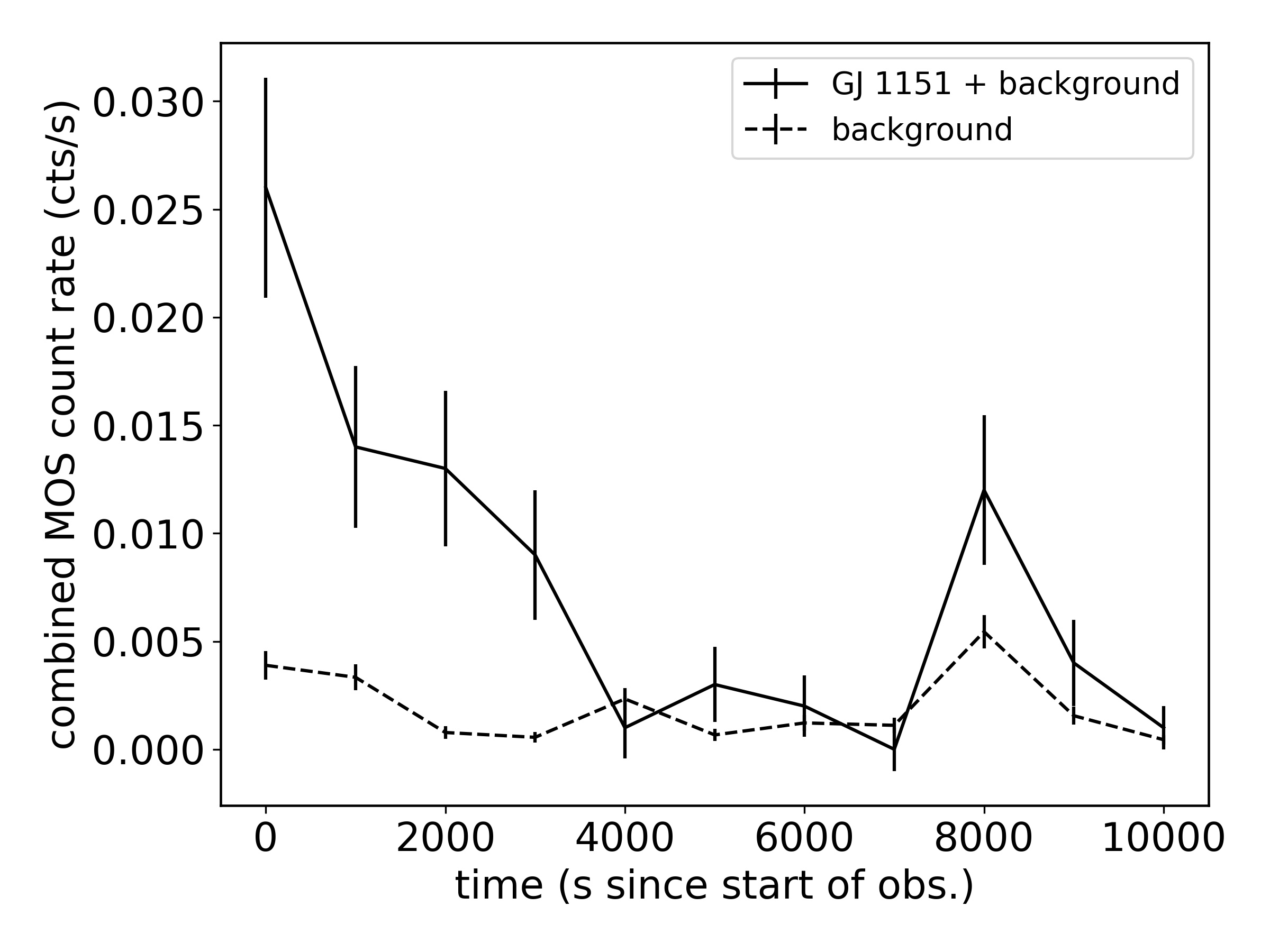}
\caption{The \textit{XMM-Newton} X-ray light curve of GJ 1151 with 1 ks time binning, using the co-added signal from both MOS detectors. The solid-line curve is the signal from the source region containing the true source signal and the underlying background, the background itself as estimated from a nearby region is shown as a dashed line. The star shows variability, possibly the decay of a flare at the beginning of the light curve.}
\label{fig:lc}
\end{figure}

We extracted light curves with a time bin size of 1000 seconds from the source and background extraction regions of the two MOS cameras. We co-added the signal from the MOS cameras, and show the signal from the source region and the background regions (scaled to the source region size) in Fig.~\ref{fig:lc}. The corona of GJ 1151 displays some variability: in the middle of the observation the stellar X-ray emission is indistinguishable from the background count rate, but at the beginning of the observation we seem to be witnessing the decay of a stellar flare. Unfortunately, the peak of the flare was not observed so that typical relations of flare decay times to the length of the flaring coronal loop \citep{Reale2007} can not be applied here. Another possibility for the shape of the light curve at the beginning of the observation is rotational modulation of the corona, with an active region rotating from the front of the star to the back. However, as GJ~1151 has a rotation period of more than 100 days, we consider this to be less likely than a flare decay.

There is also a short spike in the source signal towards the end of the observation, but since the background spikes at the same time and the source signal is compatible with the background within $2\sigma$, it is unclear if this represents another flare or not.

We note for completeness that another mechanism for coronal brightness changes is the occurrence of coronal dimmings, which are observed to take place on our Sun after flares which are accompanied by coronal mass ejections \citep{Hudson1996, Thompson1998}. However, with the X-ray data present for GJ~1151 it is not possible to distinguish between coronal quasi-quiescence versus coronal dimmings caused by coronal mass ejections.

We tried to determine the number of excess counts after the flare has been excluded in order to quantify the quiescent flux of GJ~1151. We therefore compared the counts in the source and background regions for time stamps after the first 4000 seconds of the observation which resulted in a non-detection. The corresponding $3\sigma$ upper limit to GJ 1151's count rate during this quiescent time stretch is 2.1 counts per ks for the combined MOS detectors in the 0.2-2 keV energy band.

\subsection{GJ 1151's coronal properties from X-ray spectra}

\begin{figure}
\includegraphics[width = \linewidth]{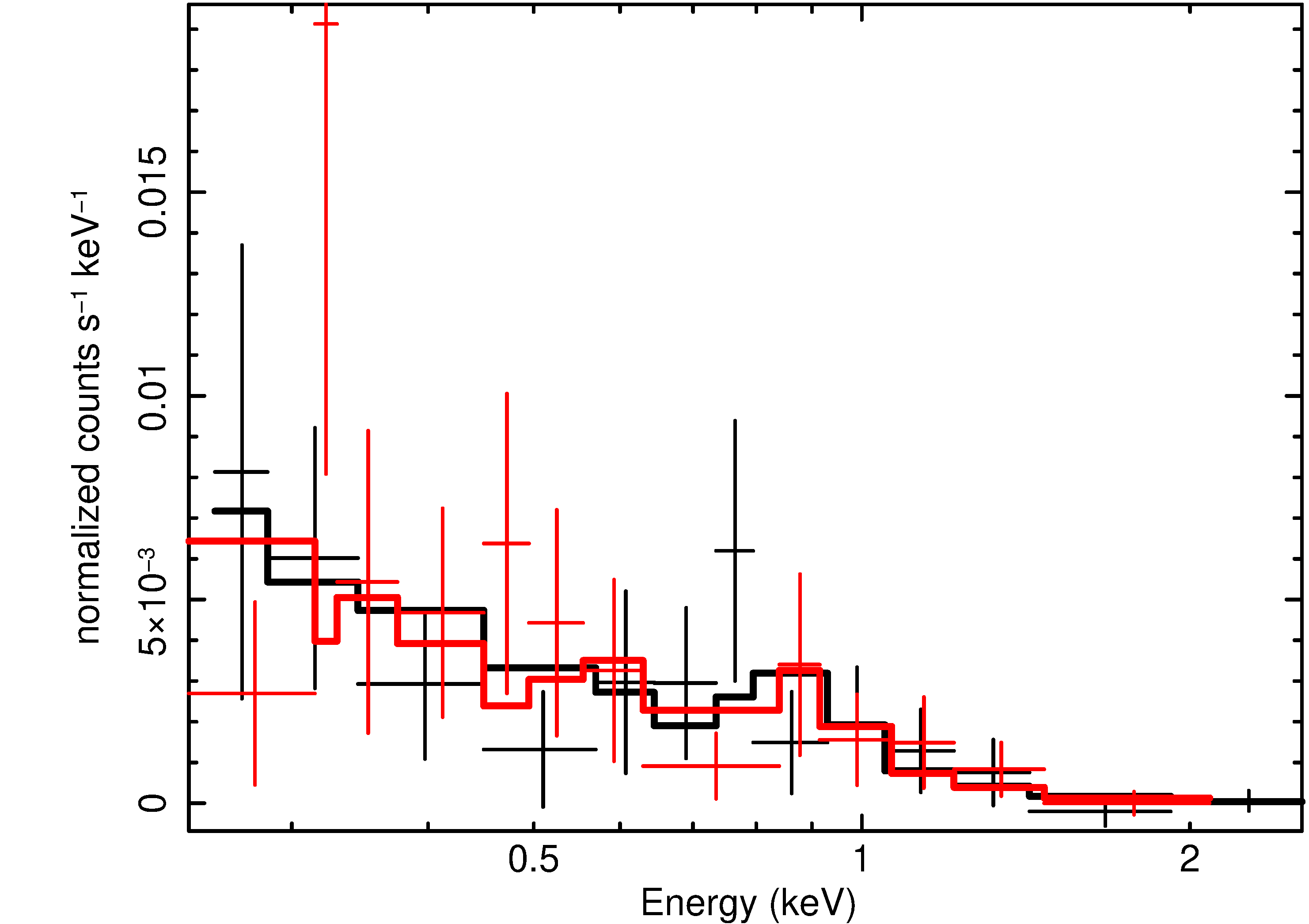}
\caption{\label{fig:extractspec} The extracted GJ 1151 spectra from \textit{XMM-Newton}'s MOS1 and MOS2 detectors (black and red crosses) are shown together with a two-temperature coronal plasma model fit (thick solid lines). The spectrum is very soft with an average coronal temperature of ca.\ 1.6~MK.}
\end{figure}


\begin{table}
\caption{\label{tab:fit} Best-fit parameters of the two-temperature coronal model to the MOS data.}

\centering
\begin{tabular}{l l}
\hline\hline
Parameter & Value \\
\hline 
k$T_1$ (keV) &  $0.095^{+0.03}_{-0.02}$  \\[0.2cm]
norm$_1$ ($\times 10^{-5}$) &  $5.2^{+7.1}_{-2.6}$  \\[0.2cm]
k$T_2$ (keV) &  $0.74^{+0.17}_{-0.25}$  \\[0.2cm]
norm$_2$ ($\times 10^{-5}$) &  $0.41^{+0.11}_{-0.09}$  \\[0.6cm]

flux (erg cm$^{-2}$ s$^{-1}$), 0.2-2 keV & $7.1^{+0.7}_{-4.6}  \times 10^{-14}$ \\[0.2cm]
flux (erg cm$^{-2}$ s$^{-1}$), 0.1-2.4       keV & $1.4^{+0.7}_{-0.9}  \times 10^{-13}$ \\[0.6cm]

$L_X$ (erg s$^{-1}$), 0.2-2 keV   & $5.5^{+0.5}_{-3.6}  \times 10^{26}$ \\[0.2cm]
$L_X$ (erg s$^{-1}$), 0.1-2.4 keV & $1.1^{+0.4}_{-0.7}  \times 10^{27}$ \\[0.1cm]

\hline
\end{tabular}
\end{table}

We extracted CCD spectra of GJ~1151 from the data of the two MOS cameras. We used Xspec version 12 to fit the spectra with an \textit{APEC} coronal plasma model \citep{Smith2001, Foster2012}, using solar-like coronal abundances from \citet{Grevesse1998}. 
Since the number of excess counts is small, we decided to group the counts into bins of at least three photons and appropriately use the Cash statistic \citep{Cash1979} for spectral fitting. A single-temperature model did not yield a satisfactory fit, with a Cash statistic value of 35.9 with 28 degrees of freedom; the single-temperature model yielded a coronal temperature of 2.9~MK, but systematically underpredicted the spectral counts at energies below 0.5~keV. Therefore we used a two-component temperature model, which yielded a Cash statistic of value of 24.4 for 26 degrees of freedom. We note here that the Cash statistic, unlike the $\chi^2$ statistic, does not yield a direct null hypothesis probability. However, the difference of the Cash statistic value between one model fit and another is distributed approximately as the difference in $\chi^2$ values for the two models, if count numbers were high enough for the $\chi^2$ statistic to be applicable. Therefore we judged that the two-temperature model, where the value of the Cash statistic is close to the number of degrees of freedom (i.e.\ similar to a reduced $\chi^2$ value of unity), is a satisfactory fit. We note that since GJ~1151 is located at a distance of only 8~pc to the Sun, spectral effects of X-ray absorption by the interstellar medium can be ignored.


We display the MOS spectra, together with the best-fit model, in Fig.~\ref{fig:extractspec}. The parameters of the best-fit model are listed in Table~\ref{tab:fit}. The emission measure (characterized by the APEC model's ''norm'' parameter) of the lower-temperature component is not very well constrained, because it is close to the lower end of the detectors' energy sensitivity and a lower fitted temperature could be offset by a larger emission measure. Still, from the spectral shape it is clear that GJ~1151 is a very soft X-ray emitter. We note here that in principle one would expect to see variation in the spectrum during the flare decay; however, the signal-to-noise of the spectrum is too low to allow such an analysis, which is why we only calculate a spectral fit for the fully time-integrated observation. 

We calculate the average coronal temperature of GJ~1151 to be $1.6$MK. The X-ray flux and luminosity is highly dependent on the energy band that is desired for this quantity, because the temperature of its corona is so low that a significant fraction of the fitted flux is located at extremely soft energies below 0.2 keV, which are not observable by XMM-Newton. We determine GJ~1151's X-ray flux and luminosity in the 0.2-2 energy band to be $F_X = 7.1\times 10^{-14}$\flux ($68\%$ confidence interval), and $L_X = 5.5 \times 10^{26}$\,erg/s. If we extrapolate the flux to an energy band of 0.1-2.4 keV, as was used by ROSAT, we find $F_X = 1.4\times 10^{-13}$\flux, and $L_X = 1.1 \times 10^{27}$\,erg/s. We note that the uncertainties on the X-ray flux and luminosity are rather large, as reported in Table~\ref{tab:fit}, due to the aforementioned uncertainty in the temperature of cooler component.

If we use the same spectral shape to quantify the upper limit of the flux during the quiescent time period, we find the upper limit to be $F_{X,\, qui} \leq 4.8\,[9.5]\, \times 10^{-14}$\flux and $L_{X,\, qui} \leq 3.7\,[7.5]\, \times 10^{26}$\,erg/s for the 0.2-2 keV [0.1-2.4 keV] energy band. It is likely that the corona of GJ~1151 is even cooler during the quiescent time, which would make the flux even lower than our upper limit.

This places GJ~1151 among low-mass stars of low magnetic activity. We estimate GJ 1151's bolometric luminosity to be $1.37\times 10^{31}$\,erg/s; we base this on GJ~1151's mass of 0.167\,$M_\odot$ as reported by \citet{Newton2017} and interpolate the bolometric luminosity from the tabulated values of \citet{Pecaut2013}\footnote{updated table values available at \url{https://www.pas.rochester.edu/~emamajek/EEM_dwarf_UBVIJHK_colors_Teff.txt}}.
Therefore GJ~1151's coronal activity indicator is $\log L_X/L_{bol} = -4.4$ in the 0.2-2 keV energy band and -4.1 in the 0.1-2.4 legacy ROSAT energy band. This places GJ~1151 towards the lower end of the activity levels displayed by the very slowly rotating low-mass stars studied by \citet{Wright2018}.

\subsection{Consistency with previous upper limits}

Two upper limits on GJ~1151's X-ray luminosity exist, one from the ROSAT All-Sky Survey (RASS) and one from a Chandra ACIS-S observation (ObsID 18944, Chandra observation cycle 18, 2.9 ks exposure time, PI Wright). 

Revisiting the Chandra observation, we find that there is actually a marginal excess of counts at the location of GJ~1151 in the 0.2-2 keV energy band, namely 3 X-ray photons in a circular region with $2^{\prime\prime}$ radius placed at the nominal position of the star, versus a background signal of 0.041 expected counts for the same region size. This corresponds to a detection at $99.7\%$ confidence level, albeit with a highly uncertain excess count measurement of $3.0^{+2.1}_{-1.4}$ counts with $1\sigma$ uncertainties, or correspondingly $1.0^{+0.7}_{-0.5}$ counts per ks \footnote{We note here that the low count numbers produce strong deviations from a Gaussian uncertainty regime. The $N\sigma$ confidence range is therefore no longer given by symmetrically multiplying the $1\sigma$ range limits by a factor of $N$.}. Since Chandra's ACIS-S detector has become less sensitive to very soft-energy photons due to a deposit accumulation on its filters, it actually traces mostly photons with energies above 0.7 keV in this observation. 

If we use our best-fit model from XMM-Newton and use the ACIS-S effective area at the time of the Chandra observation, we would expect a count rate of 4.5 counts per ks. This is higher than what is seen in the Chandra observation, which means that the star is likely not flaring during the Chandra observation. If we choose to use the same underlying spectrum as seen in XMM-Newton, the detected Chandra count number corresponds to a flux of $1.8\times 10^{-14}$\flux, which is likely an underestimate since GJ~1151's coronal X-ray emission would be even softer when the star is not flaring. This is in overall agreement with \citet{Wright2018}, who derive an upper limit from this Chandra observation for both the Chandra (0.5-8 keV) and ROSAT (0.1-2.4 keV) energy bands of $1.4\times 10^{-14}$ and $2.0\times 10^{-14}$, respectively. The small discrepancy to our detected flux seems to stem from their assumption of a coronal temperature around 0.5 keV, which is indeed often observed for fully convective M dwarfs, but is significantly lower in GJ 1151's corona as the XMM-Newton detection shows.


The RASS observation only has an accumulated exposure time of about 370~s at the position of GJ 1151, corresponding to a non-restrictive upper limit of $1.5\times 10^{-13}$\flux in the native ROSAT energy band of 0.1-2.4 keV, using our measured average coronal temperature of 1.6~MK.


\section{Discussion}

The coronal X-ray brightness of GJ~1151 is not unusual for low-activity M dwarfs. Similar X-ray activity levels have been found for slowly rotating M dwarfs by \citet{Wright2018}. However, in the case of GJ~1151 we were able to show that its corona is of a very low temperature, which means that a considerable fraction of its X-ray flux is to be found at very soft energies below 0.3 keV. 

Other slowly-rotating M dwarfs have been found to flare occasionally, see for example \citet{Raetz2020}, so the fact that GJ~1151 as a low-activity star happens to flare in the XMM-Newton observation is not extraordinary.

In the context of star-planet interactions, the coronal properties we derived for GJ~1151 from our X-ray detection do not contradict the model presented by \citet{Vedantham2020}, who based their analysis on the X-ray upper limits available at that time. Specifically, \citet{Vedantham2020} excluded radio flares as an explanation of the radio observations based on an assumed coronal temperature of 2 MK. This is very close to our measured average coronal temperature of 1.6 MK, and following the outlined calculations in \citet{Vedantham2020} a lower coronal temperature would lead to an even lower radio brightness temperature, strengthening their exclusion of radio flares as an explanation. Unfortunately, since the peak of the flare was no included in the X-ray observation, it is not possible no draw further inferences on the flare properties, such as loop length or any type of density analysis of the flaring loop.

The star-planet interaction scenario with open stellar field lines used by \citet{Vedantham2020} assumes a coronal temperature of 1 MK as the base for the stellar wind, and this can be considered realistic given our analysis. Since the X-ray observation contains a flare and the measured coronal temperature, averaged over the full observation, is 1.6 MK, one can assume that the corona of GJ~1151 is even cooler during quiescent times. The relationship of lower X-ray luminosities with lower coronal temperatures is very well established \citep{Telleschi2005, Schmitt1997, Guedel1997, Johnstone2015}. 

We note that the Poynting flux derived by \citet{Vedantham2020} of $F_P \sim 10^{23}$\,erg/s is so low that a direct detection of coronal emission induced by star-planet interaction of GJ 1151 with a nearby planet is not in the feasible range for current X-ray observatories.

\section{Conclusions}

We have detected coronal X-ray emission from the M dwarf star GJ~1151, using \textit{XMM-Newton}. The star displays coronal variability, a low coronal temperature of 1.6~MK and an average X-ray luminosity of $5.5\times$10$^{26}$ ergs s$^{-1}$ in the 0.2-2~keV energy band. The detected X-ray emission is compatible with a reported scenario of sub-Alfv\'enic star-planet interaction, motivated by the star's observed emission at radio wavelengths.

\section*{Acknowledgements}
The authors thank J.\ Callingham and H.\ Vedantham for helpful discussions.
This work is based on observations obtained with XMM-Newton, an ESA science mission with instruments and contributions directly funded by ESA Member States and NASA. Part of this work was supported by the German \emph{Leibniz-Gemeinschaft} under project number P67-2018.

\section*{Data Availability}
The data used here is publicly available in ESA's XMM-Newton data archive.




\bibliographystyle{mnras}
\bibliography{paperbib} 






\bsp	
\label{lastpage}
\end{document}